\documentclass[journal=nalefd,manuscript=article, layout=twocolumn]{achemso}
\usepackage[version=3]{mhchem} 
\usepackage{amsfonts, amsmath, amsthm}
\usepackage{caption}
\usepackage{graphicx, nicefrac}
\usepackage{dcolumn}
\usepackage{bm}
\usepackage{xr} 
\usepackage{float}
\usepackage{cancel}
\usepackage{hyperref}
\hypersetup{
    colorlinks=true,
    citecolor=black,
    urlcolor=blue,
    }

\newcommand{\ms}{1L-\ce{MoS2}\,}

\makeatletter
\newcommand*{\addFileDependency}[1]{
  \typeout{(#1)}
  \@addtofilelist{#1}
  \IfFileExists{#1}{}{\typeout{No file #1.}}
}
\makeatother

\newcommand*{\myexternaldocument}[1]{%
    \externaldocument{#1}%
    \addFileDependency{#1.tex}%
    \addFileDependency{#1.aux}%
}
\myexternaldocument{mos2_nanoletters_SI}

\author{Tristan L Britt}
\email{tristan.britt@mail.mcgill.ca}
\affiliation[McGillPhysics]{%
 Department of Physics, Center for the Physics of Materials, McGill University, 3600 rue Université, Montréal, Québec H3A 2T8, Canada}

\author{Qiuyang Li}
\affiliation[NYUChem]{Department of Chemistry, Columbia University, New York City, New York 10027, USA}

\author{Laurent P. Ren\'e de Cotret}
\affiliation[McGillPhysics]{%
 Department of Physics, Center for the Physics of Materials, McGill University, 3600 rue Université, Montréal, Québec H3A 2T8, Canada}
 
\author{Nicholas Olsen}
\affiliation[NYUChem]{Department of Chemistry, Columbia University, New York City, New York 10027, USA}

\author{Martin Otto}
\author{Syed Ali Hassan}
\affiliation[McGillPhysics]{%
 Department of Physics, Center for the Physics of Materials, McGill University, 3600 rue Université, Montréal, Québec H3A 2T8, Canada}
 
\author{Marios Zacharias}
\affiliation{
Department of Mechanical and Materials Science Engineering,
Cyprus University of Technology, P.O. Box 50329, Limassol 3603, Cyprus
}
\altaffiliation{
Institut FOTON - UMR 6082, Univ Rennes, INSA Rennes, 20 Avenue des Buttes de Coesmes, Rennes 35700, France
}
\author{Fabio Caruso}
\affiliation{
  Institut für Theoretische Physik und Astrophysik, Christian-Albrechts-Universität zu Kiel, Leibnizstraße 15,  Kiel 24118, Germany
}%

\author{Xiaoyang Zhu}
\affiliation[NYUChem]{Department of Chemistry, Columbia University, New York City, New York 10027, USA}
\author{Bradley J Siwick}%
\affiliation[McGillPhysics]{%
 Department of Physics, Center for the Physics of Materials, McGill University, 3600 rue Université, Montréal, Québec H3A 2T8, Canada}
\email{bradley.siwick@mcgill.ca}
\altaffiliation{Department of Chemistry, McGill University, 801 rue Sherbrooke Ouest, Montréal, Québec H3A 0B8, Canada}
\title[UEDS+BTE]
  {Direct view of phonon dynamics in atomically thin \ce{MoS2}}

\abbreviations{TMDC, DFT, BTE, EPC, UV, NIR}

\begin{document}

\begin{tocentry}
    \includegraphics[width=3.25in]{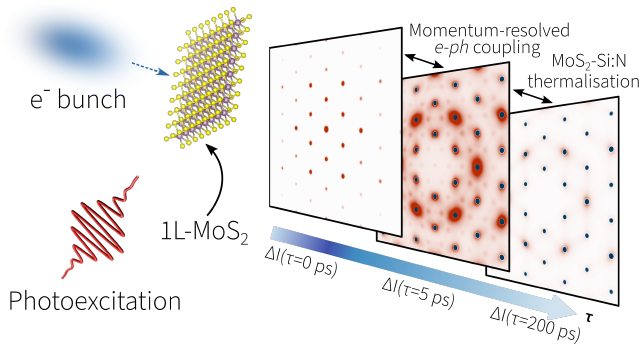}
\end{tocentry}

\begin{abstract}
    Transition metal dichalcogenide monolayers and heterostructures are highly tunable material systems that provide excellent models for physical phenomena at the two-dimensional (2D) limit. While most studies to date have focused on electrons and electron-hole pairs, phonons also play essential roles. Here, we apply ultrafast electron diffraction and diffuse scattering to directly quantify, with time and momentum resolution, electron-phonon coupling (EPC) in monolayer molybdenum disulfide (\ce{MoS2}) and phonon transport from the monolayer to a silicon nitride (\ce{Si3N4}) substrate. Optically generated hot carriers result in a profoundly anisotropic distribution of phonons in the monolayer on the $\sim$ 5 ps time scale. A quantitative comparison with \textit{ab-initio} ultrafast dynamics simulations reveals the essential role of dielectric screening in weakening EPC. Thermal transport from the monolayer to the substrate occurs with the phonon system far from equilibrium. While screening in 2D is known to strongly affect equilibrium properties, our findings extend this understanding to the dynamic regime. 
\end{abstract}


\section{\label{sec:intro}Introduction\protect}
\begin{figure*}[t!]
    \includegraphics[width=\textwidth]{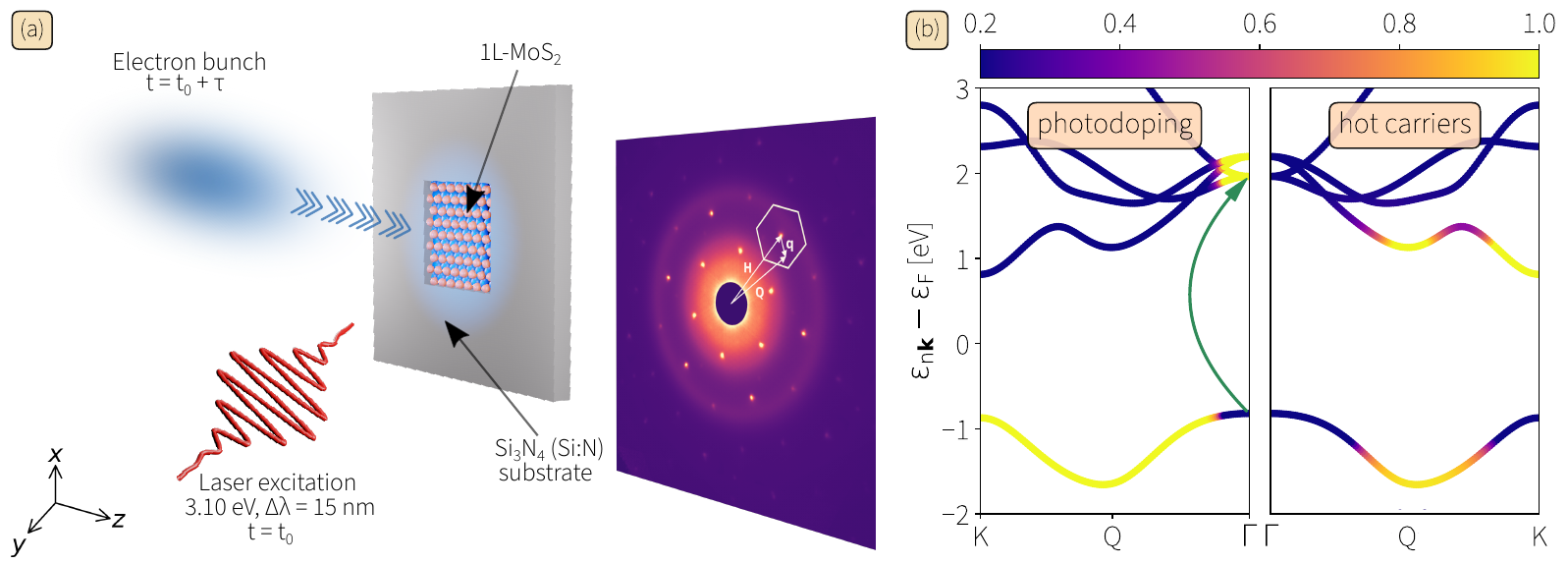}
    \caption{\label{fig:UEDS} (a) Schematic of the pump-probe UED and UEDS experiment.  The \ce{MoS2} monolayer-\ce{Si3N4} (abbreviated Si:N) heterostructure is photoexcited (pumped) by an ultrashort laser pulse (400nm, 3.1 eV, fluence of 1.8 mJ-cm$^{-2}$) at $t=t_0$. After a controllable delay time $\tau$, an ultrashort electron pulse probes the sample at normal incidence (in $\hat{z}$), and transmission elastic/inelastic (phonon-diffuse) scattering is imaged at the detector (right). A representative Brillouin zone of \ms with the Bragg peak $\mathbf{H}$, the scattering vector $\mathbf{Q}$, and reduced scattering vector $\mathbf{q}$ are indicated. (b)  Schematic of the photodoped carrier populations formed via photoexcitation (left) and after rapid carrier thermalization (right). The color bar represents electron occupation. Conduction band electrons appear yellow and valence band holes appear blue. Hot carriers in the right panel are described by a Fermi-Dirac distribution at an elevated charge carrier temperature $T_0^\text{cc}\simeq1700$K.}
\end{figure*}

2D Transition-metal dichalcogenides (TMDs)  are currently  subject of intense research due to their exciting electronic and opto-electronic properties.  These arise from the highly correlated nature of their spin\cite{Zhou2019,MolinaSanchez2013}, valley\cite{Xiao2012,Beyer2019}, electronic\cite{Qiu2013},
and vibrational\cite{Yao2008} degrees of freedom which can vary dramatically with the number of layers, most notably with the breaking of spatial inversion symmetry in monolayers. In molybdenum disulfide (\ce{MoS2}) monolayers (1L), strong spin-orbit coupling additionally \cite{Kadantsev2012} allows for direct control of the spin and valley degrees of freedom \cite{Xiao2012,Song2013} and the possibility of exciting chiral phonons \cite{Zhang2015, Zhu2018}. The coupling of monolayers in heterostructures \cite{Balch2020}, moir\'e superlattices \cite{Zhang2018, Maity2022} or to underlying substrates \cite{Trainer2020} and modified dielectric environments \cite{Borghardt2017} have opened further avenues for wide tunability in properties and control of correlated phases.  

Many effects in these (and other) 2D materials arise specifically from the interplay between electronic and vibrational degrees of freedom, known as electron-phonon coupling (EPC)\cite{Giustino2017}. EPC is a central concern in materials physics generally, providing the fundamental origin of phenomena as diverse as conventional superconductivity, charge-density waves and soft-mode phase transitions, playing an important role in determining both charge and thermal transport properties. However, unlike their 3D counterparts where EPC is a fundamental property of the materials, EPC should be uniquely sensitive to the environment of a 2D material where the Coulomb potential is known to be poorly screened within the 2D layer \cite{XYZ2015}. 

Here we show that ultrafast electron diffraction (UED) combined with the recently developed technique of ultrafast electron diffuse spectroscopy (UEDS) can provide a momentum-resolved picture of the nonequilibrium phonon dynamics that follow photocarrier generation in monolayer \ce{MoS2}.  This electron analog of ultrafast diffuse x-ray scattering \cite{trigo2013fourier, teitelbaum2021measurements} has previously provided momentum-resolved information on inelastic electron-phonon scattering \cite{Stern2018,Waldecker2017,Chase2016, Helene2021}, soft phonon physics, \cite{Otto2021} and polaron formation \cite{RenedeCotret2022} in bulk materials.  The extension of this technique to monolayers is possible due to advancements in instrumentation and the $>10^5$ enhancement in atomic scattering cross section for electrons compared to x-rays. This provides access to novel phenomena, including effects of the local dielectric environment on EPC within the monolayer.  

The technique also provides access to the microscopic, nonequilibrium phonon population distribution at all momenta during phonon (heat) transport across the monolayer-substrate interface. Comparing the experimental results with first-principles calculations of the coupled electron-phonon dynamics\cite{Caruso2021, Zacharias2021} provides validation of this approach for the same in-monolayer processes probed experimentally. Overall, the current work demonstrates a new approach to probe EPC and nanoscale phonon transport within a monolayer/substrate heterostructure that is broadly applicable to monolayer heterostructures and moir\'e superlattices. 

The opto-electronic properties, exciton physics \cite{Wang2018} and ultrafast carrier dynamics in monolayer TMDs have been extensively studied by both conventional and ultrafast spectroscopies \cite{Liu2019, Lee2021, He2020, Li2015, Kime2017}. To date, experimental investigations of EPC effects and nonequilibrium phonon dynamics in monolayer and few-layer TMDs have been limited to what can be learned about zone-centered optical phonons via conventional and time-resolved Raman spectroscopy \cite{Sohier2019, Quan2021, Ferrante2021} or averaged atomic mean-squared displacements via UED\cite{Mannebach2015,HeXing2020,Luo2021}. The momentum selectivity of phonon emission and the time-dependent anisotropic phonon populations, expected throughout the Brillouin zone (BZ) of TMDs\cite{Caruso2021,Waldecker2017}, is hidden from view with these techniques. By contrast, UEDS provides momentum-resolved information on EPC and its modification due to (i) the dielectric environment, (ii) nonequilibrium phonon relaxation in \ms (including anharmonic decay within the monolayer), and (iii) phonon transport to the underlying substrate as we show here.

\begin{figure}[t!]
    \centering
    \includegraphics[width=0.7\linewidth]{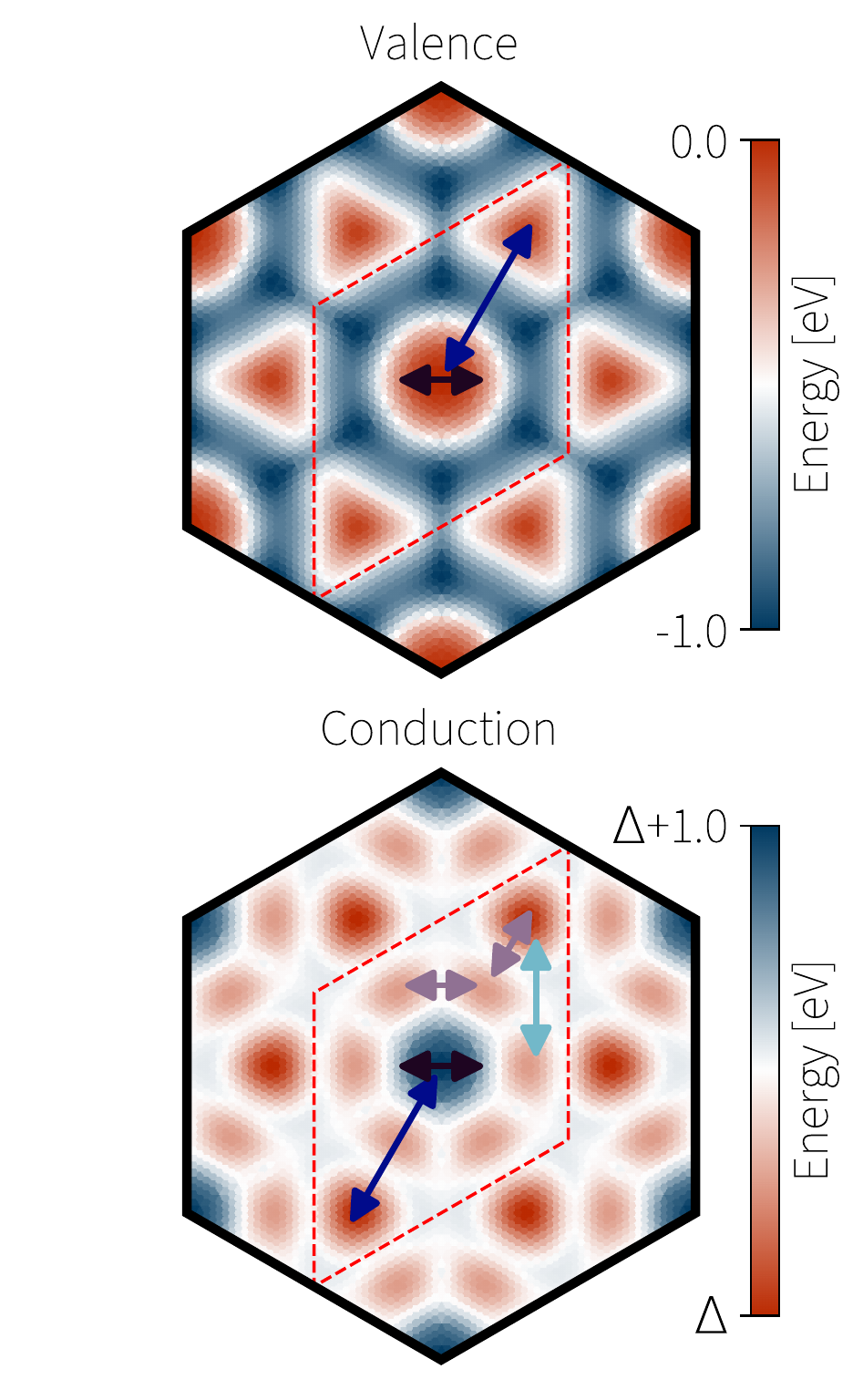}
    \caption{The structure of the valence and conduction bands of \ms indicating available intra-valley and inter-valley inelastic phonon scattering pathways for phonon-mediated carrier cooling. $\Gamma$ (black), $K$ (dark blue), $M$ (light blue), and $Q$ (purple) phonon scattering processes are shown for electron (hole) relaxation. Umklapp processes are not shown. Inset in red is the $\Gamma$-centered rhomboidal primitive unit cell in reciprocal space. For the conduction band, values are given relative to $\Delta$, the Kohn-Sham direct band-gap.}
    \label{fig:disp_band}
\end{figure}

In these pump-probe experiments (Fig \ref{fig:UEDS}a), the excitation photon energy of 3.1eV (400 nm) is resonant with vertical excitation near zone center, photodoping electrons (holes) into the conduction (valence) bands as shown in Fig \ref{fig:UEDS}b (left panel). These nonequilibrium charge carriers rapidly relax towards the band edges ($K$ and $Q$ points) in momentum and energy via electron-electron scattering \cite{Caruso2021} as indicated in Fig \ref{fig:UEDS}b (right panel). Carrier cooling proceeds primarily through inelastic electron-phonon scattering, which is expected to be strongly anisotropic in momentum due to the structure of the valence and conduction band valleys (Fig. \ref{fig:disp_band}). It is this electron-phonon coupling (from the perspective of the phonon system) that is followed with time and mode resolution with UEDS through the elastic and inelastic phonon scattered electron intensity measured as a function of scattering vector,  $I(\mathbf{Q},\tau)$ [\AA$^2$] (see Fig \ref{fig:UEDS}). See Methods for details.

\begin{figure*}[th!]
    \centering
    \includegraphics[width=\linewidth]{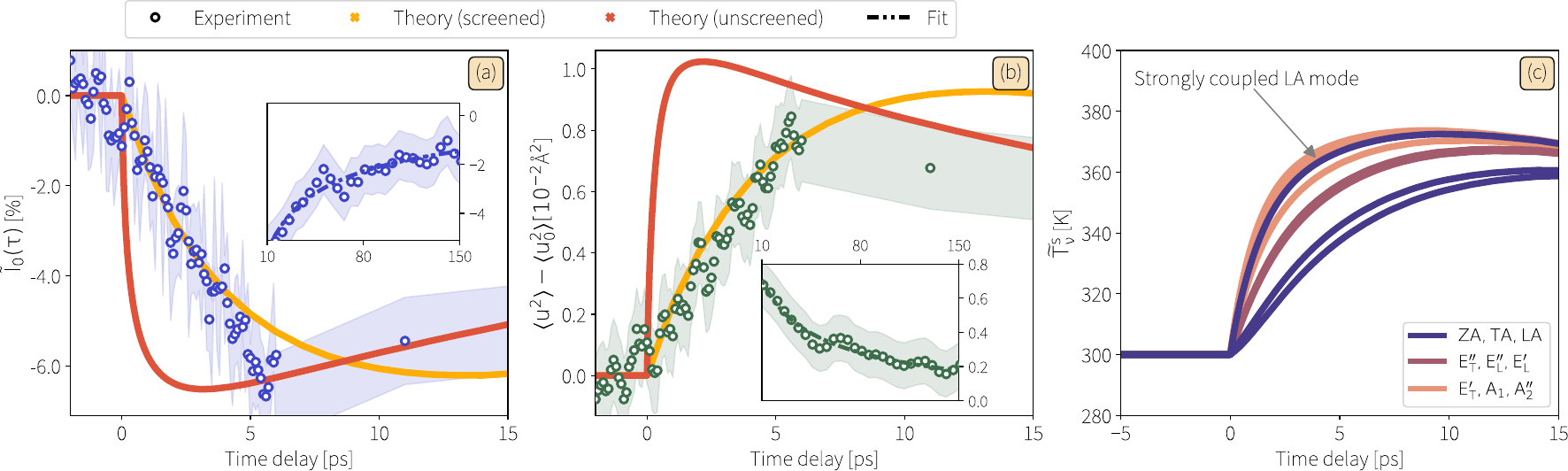}    %
    \caption{Photocarrier-phonon equilibration in \ms as measured via Bragg peak Debye-Waller dynamics. (a) Relative change in (300) Bragg peak intensity following photoexcition (blue circles).  First-principles calculation of free-standing \ms (red curve), and including the dielectric environment provided by the Si:N substrate (orange curve). The best-fit (dashed lines) are given for long times, described by the 1D heat kernel (see text). The inset highlights the long-time behavior of each signal and the blue band represents the 1$\sigma$ uncertainty bound on the data points.  (b) Increase in MSD extracted from Bragg intensity by Eqn \ref{eqn:MSD}. Red, orange, and dashed curves as in panel (a). (c) Average phonon mode temperature (momentum integrated) from \textit{ab-initio} simulations including the Si:N dielectric environment.
    }
    \label{fig:bragg+u2+temp}
\end{figure*}
\section{\label{sec:level2-bragg}Dielectric Screening of EPC in the \ms/Si:N Heterostructure}

Following photoexcitation, all measured \ms Bragg peaks are suppressed \cite{Grosso2014,Fultz2008} due to the Debye-Waller effect\cite{Debye1913, Waller1923} (Fig \ref{fig:bragg+u2+temp}a). The relative peak intensity is directly related to the increase in in-plane atomic mean-squared displacements (MSD) $\langle u^2\rangle(\tau)$ given by:
\begin{equation*}
    \langle u^2\rangle(\tau)-\langle u^2_0\rangle = 
\end{equation*}
\begin{equation}
    \label{eqn:MSD}
    -\frac{3}{4\pi^2}\frac{\ln\{\widetilde{I}_0(\mathbf{H}_{ij},\tau)/\widetilde{I}_0(\mathbf{H}_{ij},\tau<\tau_0)\}}{|\mathbf{H}_{ij}|^2}
\end{equation}
where $\langle u^2_0\rangle$ is the average of the equilibrium in-plane MSD tensor, and $\mathbf{H}_{ij}$ is the scattering vector of the \ms Bragg reflection with Miller index  ${(i,j)}$. The measured changes in MSD following photoexcitation is shown in Fig \ref{fig:bragg+u2+temp}b.

The transient rise in MSD provides an average measure of the rate at which photocarrier excitation energy is transferred to phonons in the monolayer, informing on carrier-phonon equilibration through EPC. We compare these measurements directly with the results obtained by combining together \textit{ab-initio} calculations of nonequilibrium dynamics\cite{Caruso2021} and scattering intensities\cite{Zacharias2021,Zacharias2021_PRB} (see SI) in Figs. \ref{fig:bragg+u2+temp}a-b.  The ultrafast dynamics simulations for a free-standing monolayer film are shown in red, predicting a much higher rate of MSD increase than that observed in the experimental data.  When the dielectric environment provided by Si:N is included using a semi-infinite slab model with no free parameters, there is quantitative agreement between the rise is MSD measured and that predicted within experimental uncertainties.  

The dielectric screening provided by Si:N in the heterostructure renormalizes the EPC matrix elements according to
$\tilde{g}_{mn\nu}({\bf k,q})= g_{mn\nu}({\bf k,q}) / \varepsilon^{\rm sub}_\infty$,
where $g_{mn\nu}({\bf k,q})$
is the coupling matrix element of a free-standing monolayer (see SI 3.2.3). Here, the high-frequency dielectric constant of the semi-infinite slab is the average permittivity between ambient vacuum conditions and the Si:N substrate, namely $\varepsilon^{\mathrm{sub}}_\infty=(1+\varepsilon_\infty )/2$, where $\varepsilon_\infty = 7.8$ is dielectric

constant of bulk Si:N\cite{Warlimont2005,Balland1999}.  The average value of these matrix elements for the LA and LO ($E^{'}$) phonon modes in free-standing \ms at $K$ are 19 meV and 23 meV respectively.  These are each reduced by a factor of $\varepsilon^\mathrm{sub}_\infty=4.4$ due to the presence of the Si:N dielectric environment, providing a quantitative explanation for the approximately order-of-magnitude reduction in the rate of photocarrier-phonon energy transfer compared to predictions for a free-standing film (SI, Table S1). The inclusion of the renormalized EPC matrix elements produces quantitative agreement between first-principles and experimental results in both amplitude and rate of change.
\begin{figure*}[!th]
    \centering
    \includegraphics[width=\linewidth]{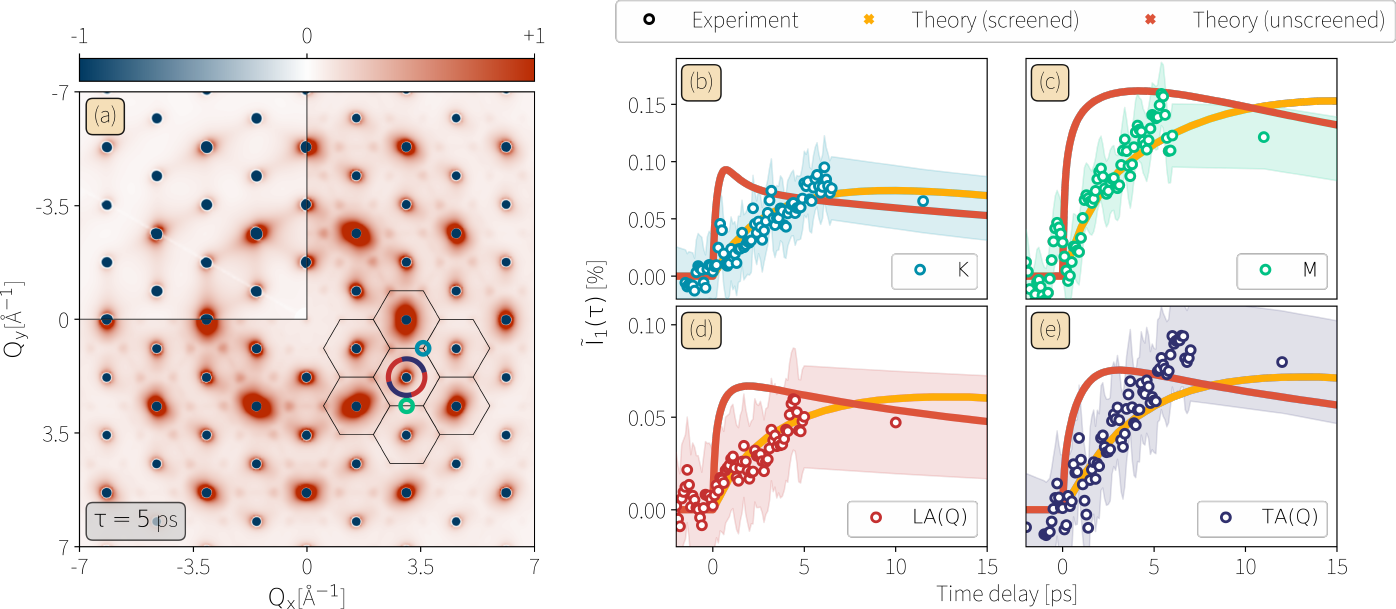}
    \caption{Momentum-resolved phonon re-equilibration dynamics (a) All-phonon differential diffuse scattering pattern of \ms calculated from first principles as $\Delta I=I(\mathbf{Q},\tau=5\text{ ps})-I(\mathbf{Q},T=300\text{ K})$. Inset (upper left) is the thermal differential diffuse scattering pattern calculated as $\Delta I = I(\mathbf{Q},T=380\text{ K})-I(\mathbf{Q},T=300\text{ K})$. The temperature of 380 K corresponds to an effective lattice temperature as extracted from the observed MSD at $\tau=5$ ps (see text), shown on the same color scale. Black hexagons indicate BZ boundaries.  Regions for which data is shown in (b-d) are indicated with the matching color. (b-d) The relative change in diffuse intensity at reduced scattering vectors (b) $K$, (c) $M$, (d) LA phonons at ${Q}$, and for (e) TA phonons at ${Q}$. Signals are obtained by integration over the colored regions in (a), as well as over every visible BZ (see SI). Acoustic signals are extracted by integrating over the segmented annuli given in (a), with LA and TA distinction possible due to phonon scattering selection rules (SI, Fig. S5). Red and orange curves as in Fig \ref{fig:bragg+u2+temp}.}
    \label{fig:sim_exp_EPC}
\end{figure*}

The excitation energy has equilibrated between carriers and phonons in \ms by $\tau\sim10$ ps, as shown by the peak value of MSD in Fig \ref{fig:bragg+u2+temp}b. The roll-over and decay of MSD for $\tau\ > 10$ ps indicates the reduction in vibrational energy in the monolayer due to phonon transport into the Si:N substrate (Fig. \ref{fig:bragg+u2+temp}c). The reported rise-time for the MSD following photoexcitation (i.e. the observed Debye-Waller decay of
Bragg peak intensities) in the case of \ms on sapphire is slower than reported here\cite{Mannebach2015,HeXing2020}, while the opposite is true for \ms on \ce{SiO2}\cite{Luo2021}. This is as expected based on the dielectric constants of these substrate materials, providing additional evidence in support of our conclusions regarding dielectric screening of the electron-phonon interaction in \ms.

Since photoexcitation was effectively uniform over the probed 250 $\mu$m region ($<$ 10\% variation), thermal transport from the photoexcited monolayer into the Si:N substrate is well described as one-dimensional (1D) on picosecond timescales; a delta-function heat impulse (following photoexcitation of the monolayer) diffuses into the substrate. In this geometry, with these initial conditions, an analytical solution for the temperature of the monolayer is given by the Green’s function of the 1D heat equation. This model provides an excellent fit of the data for long times in Fig \ref{fig:bragg+u2+temp}, far superior than a single exponential decays, yielding a thermal conductivity $\kappa=313\pm4$ Wm$^{-1}$K$^{-1}$ which is in reasonable agreement with simulated and experimental values in the $\bf{c}$-axis of Si:N\cite{Hu2020,Hirosaki2002}. We conclude the thermal boundary resistance between \ms and Si:N in the heterostructure is small. 

\section{\label{sec:level2-diffuse}Mode-Resolved Nonequilibrium Phonon Dynamics in \ms \protect}

The increase in MSD determined from the Debye-Waller suppression of Bragg peak intensities does not uniquely define the microscopic state of the phonon system. By contrast, UEDS measurements together with \textit{ab-initio} simulations yield unprecedented details of mode- and momentum-resolved nonequilibriumn phonon population distributions in the monolayer that underlie the changes in MSD observed via the Bragg peak dynamics. The transient UEDS signals from \ms following photoexcitation show the strongest increases at the $K$, $M$ and $Q$ points of the BZ (Fig \ref{fig:sim_exp_EPC}).  The time- and momentum-resolved phonon excitation dynamics at each of these points in the BZ are in good agreement with our \textit{ab-initio} predictions provided the effects of Si:N substrate dielectric screening are included (Fig \ref{fig:sim_exp_EPC}b-e).

These combined UEDS and first-principles analyses show that the nonequilibrium state of the phonon system several picoseconds after photoexcitation is profoundly anisotropic in momentum.  This anisotropy is primarily determined by the momentum-dependent electron-phonon interaction strength and the available inelastic electron-phonon scattering pathways that are open to the hot carriers. These pathways are constrained by the electronic band structure and carrier distribution (Fig \ref{fig:disp_band}) and explain the observed and computed momentum-dependent phonon heating dynamics.  Within 5 - 10 ps, the carrier and phonon systems in \ms have equilibrated with respect to the partition of excitation energy, but the phonon system remains profoundly out of equilibrium internally.

Previous work has demonstrated the possibility of defining a time-dependent effective phonon temperature, $T_\text{eff}(\tau)$, that corresponds to the observed MSD using the model\cite{Waldecker2016}
\begin{equation}
    \label{eqn:NLM}
    \langle u^2\rangle(\tau) = \frac{3\hbar}{2M}\int\limits_0^\infty\coth\bigg(\frac{\hbar\omega}{2k_BT_\text{eff}(\tau)}\bigg)\frac{F(\omega)}{\omega}d\omega
\end{equation}
where $F(\omega)$ is the phonon density of states (DOS).  However, such a $T_\text{eff}(\tau)$ provides a misleading view of the nonequilibrium state of the phonon system during carrier-phonon equilibration. This is illustrated in  Fig \ref{fig:sim_exp_EPC}a where the nonequilibrium phonon-diffuse differential scattering intensity at 5 ps is compared with a thermalized phonon-diffuse differential intensity distribution at $T_\text{eff}(5 \text{ps})$ = 380K (inset), the effective temperature determined by Eqn \ref{eqn:NLM} and the measured MSD at 5 ps (Fig \ref{fig:bragg+u2+temp}). The phonon population distribution in \ms is still profoundly nonthermal and not well described by an effective temperature.
 
\begin{figure}[!t]
    \includegraphics[width=\linewidth]{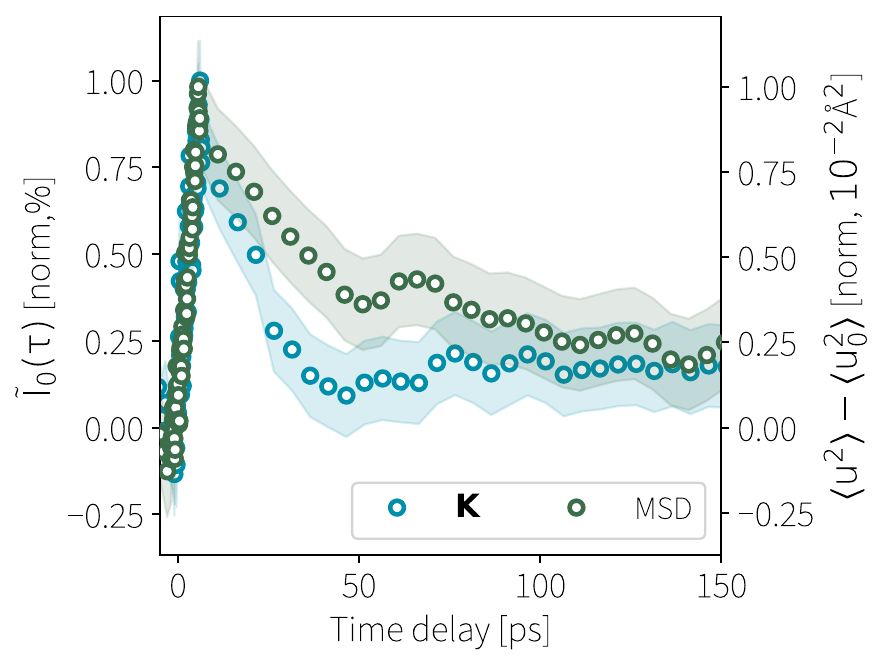}
    \caption{Optical phonon anharmonic decay vs monolayer cooling. $K$ valley optical phonon diffuse scattering (blue) compared to MSD decay (green).  The single exponential decay of the diffuse intensity at $K$ (25 ps) is in good agreement with the anharmonic decay rate of $E^{'}$ optical phonons computed via first-principles calculations (see SI Sec 3.3).}
    \label{fig:K_MSD}
\end{figure}

Further relaxation of these anisotropic nonequilibrium phonons in \ms involves coupling processes internal to the monolayer and heat transfer between the monolayer and Si:N substrate in the heterostructure.   These distinct processes are both resolved by these measurements.   In Fig \ref{fig:K_MSD}, the diffuse intensity dynamics at $K$ out to 150 ps are compared against the MSD dynamics extracted from the Bragg peaks, whose $\sim$ 50 ps decay time (single exponential fit) indicates the cooling rate of the monolayer to the underlying substrate. The observed decay of diffuse intensity at $K$ is in poor agreement with these MSD dynamics, indicating a different process is involved.  The single exponential decay time constant determined for dynamics at $K$ is 25 ps, twice as rapid as the MSD dynamics, but in good agreement with the \textit{ab-initio} anharmonic decay rate of $E^{'}$ optical phonons at $K$ (22 ps) to which UEDS is most sensitive (SI, Fig S8).   
The observed decay of the MSD is, however, in reasonable agreement with the decay in phonon-diffuse scattering measured for both the mid-BZ LA and TA modes whose heating dynamics are shown in Fig \ref{fig:sim_exp_EPC}d-e  (SI, Table S1). For times $<$ 30 ps, this sub-nanoscale phonon transport across the \ms/Si:N heterostructure interface involves a profoundly nonequilibrium population of phonons in the monolayer, including a much higher occupancy of high-wavevector acoustic phonons than would be expected of a thermalized distribution. These conditions are the result of phonon (heat) transport across the monolayer-substrate interface occurring on similar timescales as the nonequilibrium phonon relaxation within the monolayer itself.  UEDS provides a direct, momentum-resolved window on sub-nanoscale phonon transport in this far-from-equilibrium regime.   

\section{\label{sec:concl}Conclusion}
These results provide a time- and momentum-resolved view of electron-phonon coupling, anharmonic phonon decay and thermal transport in a monolayer 2D semiconductor supported on a dielectric substrate. We show that the dielectric environment provided by Si:N leads to strong renormalization of the electron-phonon coupling in the monolayer. This finding fills a critical gap in our understanding of how poor Coulomb screening at the 2D limit results in a range of physical properties sensitive to the local dielectric environment. First-principles ultrafast dynamics simulations using a recently developed framework are in excellent agreement with these measurements. Combined, these approaches provide a momentum-resolved protocol which can yield details of coupling dynamics in 2D material systems and their heterostructures.
\section{Methods}
The total scattered intensity can be decomposed into
\begin{equation}
    I_\text{all}(\mathbf{Q},\tau) = I_0(\mathbf{Q},\tau) + I_1(\mathbf{Q},\tau) + \cdots 
\end{equation}
The zeroth-order term $I_0(\mathbf{Q},\tau)$ is the elastic Bragg scattering and the first-order contribution $I_1(\mathbf{Q},\tau)$ is the inelastic single-phonon diffuse scattering that is the primary focus of this work. Adopting phonon normal mode coordinates gives:
\begin{equation}
    \label{eqn:diffuseI}
    I_1(\mathbf{Q},t) \propto\sum_{\nu}\underbrace{\frac{n_{\nu}(\mathbf{q},t)+1/2}{\omega_{\nu}(\mathbf{q},t)}}_{    
        \textstyle
        \begin{array}{c}
          |a_{\mathbf{q}\nu}|^2\\
        \end{array} 
    }
    \big|\mathcal{F}_{1\nu}(\mathbf{Q},t )\big|^2
\end{equation}
where the label $\nu$ indicates the specific phonon mode, $\bf{Q}$ is the electron scattering vector, $\bf{q}$ is the reduced phonon wavevector (i.e. $\bf{q}$ = $\bf{Q}$ - $\bf{H}$, where $\bf{H}$ is the closest Bragg peak), $a_{\bf{q}\nu}$ is the vibrational amplitude of mode $\nu$, $n_{\nu}$ is the mode-resolved occupancy with energy $\hbar\omega_\nu$, and $\mathcal{F}_{1\nu}$ are known as the one-phonon structure factors. $I_1$ provides momentum-resolved information on the nonequilibrium distribution of phonons across the entire BZ, since it depends only on phonon modes with wavevector $\bf{q}$ (Fig \ref{fig:UEDS}a). The $\mathcal{F}_{1\nu}$ are geometrical weights that describe the relative strength of scattering from different phonon modes and depend sensitively on the atomic polarization vectors $\left\{ \bf{e}_{\bf{q}\nu\kappa}\right\}$\cite{RenedeCotret2019}. Most importantly, $\mathcal{F}_{1\nu}\left(\bf{Q} \right)$ are relatively large when the phonon mode $\nu$ is polarized parallel to the reduced scattering vector $\bf{q}$.  These phonon scattering selection rules mean that $\mathcal{F}_{1\nu}$ for the out-of-plane (Z-polarized) modes and the optical modes of $E^{''}$ symmetry are very weak in the geometry of these experiments (SI, Fig S8).  These experiments primarily probe the $\mathbf{q}$-dependent population dynamics of the $E^{'}$ optical and LA/TA modes.  Terms of higher-order than $I_1$ represent multi-phonon scattering which have lower cross-sections and do not contribute significantly to the interpretation of the 1L-\ce{MoS2} signals reported here\cite{Zacharias2021, Zacharias2021_PRB}. 
The \ms/Si:N specimens used in these experiments provide two distinct contributions to $I_\text{all}$ that are both evident in Fig \ref{fig:UEDS}: (i) elastic scattering from the amorphous Si:N substrate layer which is distributed as diffuse rings, and (ii) the Bragg and phonon-diffuse scattering from the \ms.  The qualitatively different character of these signals makes the amorphous Si:N contribution to the scattering signals, $I^\text{sub}(\mathbf{Q},\tau)=I^\text{sub}(|\mathbf{Q}|)$, easily subtracted from the data as a background. See SI for details.

 The techniques of Liu \textit{et al} were used to generate the \ms sample onto the supporting Si:N substrate \cite{Liu2020}. A 150 nm-thick Au film was deposited onto a Si wafer (from Nova Electronic Materials) with e-beam evaporation (0.05 nm/s). Polyvinylpyrrolidone (PVP) solution (from Sigma Aldrich, mw 40000, 10\% wt in ethanol/acetonitrile wt 1/1) was spin-coated on the top of the Au film (1500 rpm, acceleration 500 rpm/s, 2 min) and then heated at 150 °C for 5 min. A piece of thermal release tape was used to pick up PVP/Au film and was then gently pressed onto a freshly cleaved bulk \ce{MoS2} single crystal (from HQ graphene). The thermal release tape was gently lifted up and a \ce{MoS2} monolayer is left, attached on the Au. The \ce{MoS2} monolayer was then transferred by pressing the PVP/Au film with \ce{MoS2} monolayer onto a 30 nm thick amorphous silicon nitrite (Si:N) TEM window (Norcada). The thermal release tape was then removed by heating at 130 °C. The PVP layer was removed by dissolving in deionized (DI) water for 2 h. The monolayer covered by Au was rinsed with acetone and cleaned by \ce{O2} plasma for 3 min. After removing the Au cover with a gold etchant solution (2.5g \ce{I2} and 10g KI in 100 mL DI water), the monolayer on substrate was rinsed with DI water and isopropanol, and dried under \ce{N2} flow. 
 
Electronic and vibrational properties of monolayer \ce{MoS2} have been obtained from density functional and density functional perturbation theory, respectively, and were used to calculate electron-phonon coupling matrix elements via the \texttt{Quantum Espresso}\cite{Giannozzi2017} and \texttt{EPW}\cite{Ponce2016} codes. From here, the time-propagation of the time-dependent Boltzmann equation (see SI Sec 3.2.1) yields time-, momentum-, and phonon mode-resolved occupations, which completely describe the nonthermal vibrational state. These occupations are used to compute the dynamic all-phonon diffuse scattering intensity (including Bragg scattering) within the Laval-Born-James theory\cite{Zacharias2021} as implemented in \texttt{EPW} (see SI Sec 3.2.2).

\begin{acknowledgement}

    This work was supported by the Natural Sciences and
    Engineering Research Council of Canada (NSERC), the Fonds
    de Recherche du Québec–Nature et Technologies (FRQNT),
    the Canada Foundation for Innovation (CFI) and McGill Fessenden Professorship.
    
   B.J.S. and X.Y.Z conceived the research. XYZ acknowledges support for sample preparation by the Materials Science and Engineering Research Center (MRSEC) through US-NSF grant DMR-2011738. N.O and Taketo Handa, supervised by X.Y.Z, performed PL measurements verifying the number of layers of the sample. T.L.B. performed the experiments with assistance from L.P.R.de C., M.O., S.A.H., and computed the phonon dispersion, equilibrium one-phonon structure factors, and electronic band structure. The samples were prepared by Q.L. First-principles calculations based on the TDBE were performed by F.C., who acknowledges funding from the Deutsche Forschungsgemeinschaft (DFG) - Projektnummer 443988403. The dynamic diffuse scattering intensities and MSD were computed by M.Z., who acknowledges financial support from the program META$\Delta$I$\Delta$AKT$\Omega$P of the Cyprus University of Technology. We acknowledge that the results of this research have been achieved using the DECI resource Prometheus at \href{https://www.cyfronet.pl/}{CYFRONET} in Poland with support from the PRACE aisbl.

\end{acknowledgement}

\begin{suppinfo}

    Experimental procedures, sample characterization, data preparation, and \textit{ab-initio} methodology are given in the supplementary information. Raw data is available upon reasonable request with permission from the author(s).

\end{suppinfo}

\bibliography{mos2.bib}

\end{document}